\documentclass[11pt,a4paper]{article}
\usepackage{jcappub}

\usepackage{multirow}
\usepackage{amsmath}
\usepackage{amsfonts}
\usepackage{amssymb}
\usepackage{epsfig}
\usepackage{graphicx}
\usepackage{array}

\newcommand{\pu}{\emph{Phys.\ Usp. }}
\newcommand{\aas}{\emph{Astron.\ Astrophys. }}
\newcommand{\jhep}{\emph{JHEP }}
\newcommand{\jcap}{\emph{JCAP }}
\newcommand{\prd}{\emph{Phys.\ Rev.} {\bf D }}
\newcommand{\prl}{\emph{Phys.\ Rev.\ Lett. }}
\newcommand{\plb}{\emph{Phys.\ Lett.} {\bf B }}
\newcommand{\app}{\emph{Astropart.\ Phys. }}
\newcommand{\aj}{\emph{Astrophys.\ J. }}
\newcommand{\ajl}{\emph{Astrophys.\ J.\ Lett. }}

\newcommand{\jpcs}{\emph{J.\ Phys.\ Conf.\ Ser. }}

\newcommand{\ijmpa}{\emph{Int.\ J.\ Mod.\ Phys.} {\bf A }}
\newcommand{\ijmpd}{\emph{Int.\ J.\ Mod.\ Phys.} {\bf D }}
\newcommand{\mnras}{\emph{Mon.\ Not.\ Roy.\ Astron.\ Soc. }}
\newcommand{\npps}{\emph{Nucl.\ Phys.\ Proc.\ Suppl. }}
\newcommand{\az}{\emph{Astronomic.\ Z. }}
\newcommand{\sa}{\emph{Soviet Astron. }}
\newcommand{\njp}{\emph{New J.\ Phys. }}
\newcommand{\ibid}{\emph{ibid }}

\title{IceCube expectations for two high-energy neutrino production models at active galactic nuclei}

\author[a]{C.A.~Arg\"{u}elles,}
\author[a,b]{M.~Bustamante,}
\author[a]{and A.M.~Gago}

\affiliation[a]
{Secci\'on F\'isica, Departamento de Ciencias, Pontificia
Universidad Cat\'olica del Per\'u, \\
Apartado 1761, Lima, Peru}
\affiliation[b]
{Theoretical Physics Department, Fermi National Accelerator Laboratory,\\
P.O. Box 500, Batavia, IL 60510, USA}

\emailAdd{c.arguelles@pucp.edu.pe}
\emailAdd{mbustamante@pucp.edu.pe}
\emailAdd{agago@pucp.edu.pe}

\abstract{We have determined the currently allowed regions of the parameter spaces of two representative models of diffuse neutrino flux from active galactic nuclei (AGN): one by Koers \& Tinyakov (KT) and another by Becker \& Biermann (BB). Our observable has been the number of upgoing muon-neutrinos expected in the 86-string IceCube detector, after $5$ years of exposure, in the range $10^5 \le E_\nu/\text{GeV} \le 10^8$. We have used the latest estimated discovery potential of the IceCube-86 array at the $5\sigma$ level to determine the lower boundary of the regions, while for the upper boundary we have used either the AMANDA upper bound on the neutrino flux or the more recent preliminary upper bound given by the half-completed IceCube-40 array (IC40). We have varied the spectral index of the proposed power-law fluxes, $\alpha$, and two parameters of the BB model: the ratio between the boost factors of neutrinos and cosmic rays, $\Gamma_\nu/\Gamma_\text{CR}$, and the maximum redshift of the sources that contribute to the cosmic-ray flux, $z_\text{CR}^{\max}$. For the KT model, we have considered two scenarios: one in which the number density of AGN does not evolve with redshift and another in which it evolves strongly, following the star formation rate. Using the IC40 upper bound, we have found that the models are visible in IceCube-86 only inside very thin strips of parameter space and that both of them are discarded at the preferred value of $\alpha = 2.7$ obtained from fits to cosmic-ray data. Lower values of $\alpha$, notably the values $2.0$ and $2.3$ proposed in the literature, fare better. In addition, we have analysed the capacity of IceCube-86 to discriminate between the models within the small regions of parameter space where both of them give testable predictions. Within these regions, discrimination at the $5\sigma$ level or more is guaranteed.
}

\keywords{neutrino experiments, ultra high energy photons and neutrinos, active galactic nuclei, neutrino astronomy}

\arxivnumber{1008.1396}

\begin{document}

\maketitle

\section{Introduction}\label{Section_Introduction}

Active Galactic Nuclei (AGN) are the most luminous persistent objects in the Universe, emitting radiation along almost the entire electromagnetic spectrum, with typical luminosities on the order of $10^{42}$ erg s$^{-1}$ (see, e.g., \cite{Bentz:2008rt,Aird:2009sg}). There is evidence that supports the idea that AGN are powered by matter accreting onto a central supermassive black hole, with a mass between $10^6$ and $10^{10}$ times the solar mass \cite{McKernan:2010cy,Gaskell:2009nj}. In some cases an enormous amount of energy is released in the form of two highly-collimated relativistic jets that emerge in opposite directions, perpendicularly to the accretion disc. Although the composition of these jets is unknown, it is widely believed that they contain high-energy charged particles, such as electrons, protons, and ionised nuclei, which have been accelerated as a result of the repeated crossings of the shock fronts that exist within gas clouds moving at relativistic speeds along the jets. Such a process would be able to give protons and nuclei energies of up to $\sim 10^{20}$ eV \cite{Ptitsyna:2008zs,Kachelriess:2008ze}. 
  
Recently, the Pierre Auger Observatory (PAO) claimed to have detected 69 cosmic-ray events with energies above 55 EeV \cite{:2010zzj} (see also \cite{Abraham:2007si}), providing evidence of the anisotropy in the arrival directions of utrahigh-energy cosmic rays (UHECRs). Based on the observation of 29 of these events having an angular separation of less than $3.1^\circ$ from the positions of AGN in the 12th edition V\'eron-Cetty \& V\'eron catalogue \cite{Veron:2006}, a possible correlation was found with AGN lying relatively close, at distances of 75 Mpc or less. Even though the claim on the correlation has lost some ground since the first publication of the Auger results \cite{Koers:2008ba,Farrar:2009ed}, it still constitutes a possible hint towards identifying AGN as the sources of the highest-energy cosmic rays. It is also believed that AGN could be sites of ultra-high-energy (UHE) neutrino production. These would be produced in the interactions of UHE charged particles among themselves and with the ambient photons. Therefore, under the assumption that cosmic-ray emission is accompanied by neutrino emission \cite{Waxman:1998yy,Bahcall:1999yr,Kachelriess:2010zz}, Auger's claim can be used to normalise the neutrino flux predicted by astrophysical models of AGN. 
 
In the present work, we have focused on two such models of neutrino production that take into account Auger's results: one by H.~B.~J.~Koers \& P.~Tinyakov \cite{Koers:2008hv} and another one by J.~Becker \& P.~L.~Biermann \cite{Becker:2008nf}, which we will call hereafter the KT and BB models, respectively. They differ greatly in their assumptions and, within some regions of their parameter spaces, on their predictions of the neutrino fluxes. We have assessed the possibility of observing these two fluxes in the km-scale IceCube neutrino telescope at the South Pole, by allowing their respective model parameters to vary within given boundaries, and calculating the corresponding number of high-energy muon-neutrinos expected in the detector. In doing this, we have taken into account the experimental upper bound on the neutrino flux set by the AMANDA-II experiment \cite{Xu:2008zzc}, an upper bound set by IceCube \cite{Halzen:2009tz} in its 40-string configuration, and the signal discovery potential of high-energy astrophysical neutrinos in the completed 86-string IceCube array. Furthermore, we have also explored the parameter space for regions where the event-number predictions from the two production models can be distinguished from each other.

The remaining of the paper is divided as follows. In Section \ref{Section_Theory}, we describe the salient features of the KT and BB models, and show explicitly how the observations from the PAO enter the flux normalisation. Section \ref{Section_Bounds} introduces current and envisioned experimental bounds on the high-energy extra-terrestrial neutrino flux. In Section \ref{Section_NumEvts} we allow the parameters in the KT and BB models to vary within given bounds, and calculate the number of muon-neutrinos in IceCube predicted by each, while, in Section \ref{Section_ModelCompare}, we present comparative plots of the two models in parameter space. We summarise and conclude in Section \ref{Section_Conclusions}.

\section{Two models of neutrino production at AGN}\label{Section_Theory}

AGN have long been presumed to be sites of high-energy neutrino production. In the scenario of neutrino production by meson decay, it is assumed that inside the AGN protons are accelerated through first-order Fermi shock acceleration \cite{Kachelriess:2005xh,Kachelriess:2008ze} and that pions are produced in the processes
\begin{equation}\label{EqPionProd}
 p + \gamma \rightarrow \Delta^+ \rightarrow
 \left\{\begin{array}{l}
 p + \pi^0 \\
 n + \pi^+
 \end{array}\right. ~~~~, ~~~~~
 n + \gamma \rightarrow p + \pi^- ~,
\end{equation}
with branching ratios $\text{Br}\left(\Delta^+ \rightarrow p\pi^0\right) = 2/3$ and $\text{Br}\left(\Delta^+ \rightarrow n\pi^+\right) = 1/3$. The neutral pions decay into gamma rays through $\pi^0 \rightarrow \gamma \gamma$, while the charged pions decay into electron- and muon-neutrinos through
\begin{equation}\label{EqPionDecay}
 \pi^+ \rightarrow \nu_\mu + \mu^+ \rightarrow \nu_\mu + e^+ + \nu_e + \overline{\nu}_\mu ~~~~, ~~~~~
 \pi^- \rightarrow \overline{\nu}_\mu + \mu^- \rightarrow \nu_\mu + e^- + \overline{\nu}_e + \nu_\mu ~.
\end{equation}
The gamma rays thus created may be obscured and dispersed by the medium, and the protons will in addition be deviated by extragalactic magnetic fields on their journey to Earth. Neutrinos, on the other hand, escape from the production site virtually unaffected by interactions with the medium, so that, if their direction could be reconstructed at detection, they could point back to their sources. 

If neutrinos are produced by charged pion decay, then, from eq.~(\ref{EqPionDecay}), the ratios of the different flavours ($\nu_x + \overline{\nu}_x$) to the total flux are 
\begin{equation}\label{EqRatiosProd}
 \phi_{\nu_e}^0 : \phi_{\nu_\mu}^0 : \phi_{\nu_\tau}^0 = 1 : 2 : 0 ~.
\end{equation}
Under this assumption, by the time neutrinos reach Earth, standard mass-driven neutrino oscillations will have distributed the total flux equally among the three flavours so that, at detection,
\begin{equation}\label{EqRatiosDet}
 \phi_{\nu_e} : \phi_{\nu_\mu} : \phi_{\nu_\tau} = 1 : 1 : 1 ~.
\end{equation}
New physics effects, such as neutrino decay \cite{Beacom:2003nh}, decoherence \cite{Bhattacharya:2010xj}, or violation of Lorentz invariance or of CPT \cite{Barenboim:2003jm,Bazo:2009en,Bustamante:2010nq}, could in principle result in large deviations from these ratios. In the present work, we have assumed that the ratios at production and detection are given, respectively, by their standard values, eqs.~(\ref{EqRatiosProd}) and (\ref{EqRatiosDet}).

In what follows, we will present in detail two representative models of UHE neutrino production at AGN, one by Koers \& Tinyakov (KT) and the other by Becker \& Biermann (BB), both of which make use of the apparent correlation between the directions of UHECRs and the positions of known AGN reported by the PAO in order to extrapolate the diffuse neutrino flux.

\subsection{Cosmic ray flux normalisation} 
 
The preferred mechanism for cosmic-ray acceleration at AGN is first-order Fermi acceleration \cite{Kachelriess:2008ze}, which results in a power-law differential diffuse cosmic-ray proton spectrum, 
\begin{equation}\label{EqProtonSpectrum}
 \phi_p^\text{diff}\left(E\right) \equiv \frac{dN_p}{dE} = A_p^\text{diff} E^{-\alpha_p} ~,
\end{equation}
with $E$ the cosmic-ray energy at detection on Earth and $A_p^\text{diff}$ an energy-independent normalisation constant. The integral of this expression,
\begin{equation}\label{EqIntFlux}
 \Phi_p^\text{diff}\left(E_{\text{th}}\right) 
 = \int_{E_{\text{th}}} \frac{dN_p}{dE} dE
 \simeq A_p^\text{diff} \left(\alpha_p-1\right)^{-1} E_{\text{th}}^{-\alpha_p+1} ~,
\end{equation}
is the integrated cosmic ray flux above a certain threshold energy $E_{\text{th}}$. Using experimental data, the integrated flux can also be calculated as
\begin{equation}
 \Phi_p^{\text{diff}} \left( E_{\text{th}} \right) = N_{\text{evts}} \left(E_{\text{th}}\right) / \Xi ~,
\end{equation}
where $N_{\text{evts}}\left(E_{\text{th}}\right)$ is the number of observed cosmic rays above a given value of $E_{\text{th}}$ and $\Xi$ is the total detector exposure. 

Combining this expression with eq.~(\ref{EqIntFlux}) yields for the normalisation constant,
\begin{equation}\label{EqNormConstDiff}
 A_p^\text{diff} 
 = \Phi_p^\text{diff} \left(\alpha_p-1\right) E_\text{th}^{\alpha_p-1}
 = \frac{N_{\text{evts}}\left(\alpha_p-1\right)}{\Xi} E_{\text{th}}^{\alpha_p-1} ~.
\end{equation}
We will see in the following two subsections that the relation between the cosmic-ray normalisation constant, $A_p^\text{diff}$, and the neutrino normalisation constant, $A_\nu^\text{diff}$, is model-dependent.

When calculating the proton spectrum from a single point source, we will need to weigh the normalisation constant using the detector effective area $A$ that is accessible to the observation, which depends on the declination $\delta_s$ of the source, i.e.,
\begin{equation}\label{EqNormConstPt}
 A_p^{\text{pt}} 
 = \Phi_p^{\text{pt}} \left(\alpha_p-1\right) E_{\text{th}}^{\alpha_p-1}
 \equiv \frac{N_{\text{evts}}\left(\alpha_p-1\right)}{\Xi} E_{\text{th}}^{\alpha_p-1} 
 \frac{\int A\left(\delta_s\right) d\Omega}{A\left(\delta_s\right)} ~,
\end{equation}
where we have implicitly defined the integrated flux from a point source, $\Phi_p^\text{pt}$.

We will use the latest results from the PAO on the observation of UHECRs \cite{:2010zzj} to evaluate the diffuse and point-source cosmic-ray fluxes. Using data recorded from 1 January 2004 to 31 December 2009, amounting to an exposure of $\Xi = 20 370$ km$^2$ yr sr, the total number of UHECRs with zenith angles $\theta \leq 60^\circ$ and reconstructed energies above $E_\text{th} = 55$ EeV is $N_\text{tot} = 69$ events. Of these, the arrival directions of $N_\text{corr} = 29$ events were found to lie at an angular distance of less than $3.1^\circ$ from the position of an AGN within 75 Mpc ($z \leq 0.018$) in the 12th edition V\'eron-Cetty \& V\'eron (VCV) catalogue, i.e., they were correlated to an identified AGN. In particular, $N_\text{Cen A} = 2$ events were correlated to Centaurus A (Cen A), the nearest active galaxy, which, at a distance of about 3.5 Mpc, is one of the most promising UHE neutrino sources \cite{Kachelriess:2008qx,Kachelriess:2009wn}.

Note that the original PAO report on UHECR anisotropy \cite{Abraham:2007si} made use of $9000$ km$^2$ yr sr to report a total of $29$ events above a threshold energy of $57$ EeV, out of which $20$ were correlated to AGN in the VCV catalogue, and $2$ were correlated to Cen A. The neutrino production models that we have considered in our analysis were built using these data. In what follows, we have updated them using the latest PAO results. 

\subsection{Model by Koers \& Tinyakov}\label{Section_Theory_Sub_KT}

The KT model \cite{Koers:2008hv} assumes that Cen A is a typical source of UHECRs and neutrinos, and computes the diffuse flux under the assumption that all sources are identical to Cen A by integrating over a cosmological distribution of sources, while taking into account energy losses during the propagation of the particles. Two limiting cases have been considered regarding the source distribution: one in which there is no source evolution with redshift, that is, $\epsilon\left(z\right) = 1$, and another one, adopted from \cite{Boyle:1997sm}, in which there is a strong source evolution that follows the star formation rate, i.e.,
\begin{equation}
 \epsilon\left(z\right) \varpropto 
 \left\{\begin{array}{ll}
  \left( 1 + z \right)^{3.4}   & , ~\text{if}~ z \le 1.9 \\
  \left( 1 + 1.9 \right)^{3.4} & , ~\text{if}~ 1.9 < z < 3 \\
  \left( z - 3 \right)^{-0.33} & , ~\text{if}~ z \ge 3
 \end{array}\right. ~.
\end{equation}

The integrated UHECR diffuse flux and the integrated flux from Cen A above 55 EeV can be calculated, respectively, using eqs.~(\ref{EqNormConstDiff}) and (\ref{EqNormConstPt}):
\begin{eqnarray}
 \label{EqPhiPDiff}
 \Phi_p^{\text{diff}} \left(E_\text{th}\right) 
 &=& \frac{N_{\text{tot}}-N_{\text{Cen A}}}{\Xi}
 = 1 \times 10^{-20} ~\text{cm}^{-2} ~\text{s}^{-1} ~\text{sr}^{-1} \\
 \Phi_p^{\text{Cen A}} \left(E_\text{th}\right) 
 &=& \frac{N_{\text{Cen A}}}{\Xi} \frac{\int A\left(\delta_s\right)d\Omega}{A\left(\delta_s\right)}
 = 2 \times 10^{-21} ~\text{cm}^{-2} ~\text{s}^{-1} ~,
\end{eqnarray}
where $\delta_s = -43^\circ$ is the declination of Cen A. The relative exposure at 
this declination is $A\left(\delta_s\right) / \int A\left(\delta_s\right) d\Omega = 0.15$ sr$^{-1}$ \cite{Sommers:2000us,Koers:2008hv}. In eq.~(\ref{EqPhiPDiff}), the number of cosmic rays from Cen A is subtracted from the total since the flux is not subject to the energy losses that the diffuse flux is, on account of its being the closest AGN.

The diffuse neutrino flux is normalised using the integrated UHECR flux $\Phi_p^{\text{diff}} \left(E_\text{th}\right)$ above the threshold $E_{\text{th}}$, 
\begin{equation}\label{EqScalingKT}
 \frac{\phi_\nu^{\text{diff}}\left(E_\nu\right)}{\phi_\nu^{\text{Cen A}}\left(E_\nu\right)}
 = H \left(E_\text{th}\right) \frac{\Phi_p^{\text{diff}}\left(E_{\text{th}}\right)}{\Phi_p^{\text{Cen A}}\left(E_{\text{th}}\right)} 
 \simeq 5 H \left(E_\text{th}\right) ~.
\end{equation}
The proportionality constant, $H\left(E_{\text{th}}\right)$, is called the ``neutrino boost factor'' and contains the information on neutrino mean free path lengths and source evolution. To calculate it, proton energy losses are taken into account in the continuous-loss approximation, considering losses by the adiabatic expansion of the Universe and from interactions with the CMB photons resulting in pion photoproduction and electron-positron pair production; see Appendix A in Ref.~\cite{Koers:2008hv} for details. The variation of $H$ with $\alpha_p$ is shown in the same reference. Note that the change in the reconstructed threshold energy from $57$ EeV in the original PAO analysis \cite{Abraham:2007si} to $55$ EeV in the updated analysis \cite{:2010zzj} has reduced $H$ in about $10\%$. This decrease is compensated by a higher value of the ratio $\Phi_p^\text{diff}/\Phi_p^\text{Cen A}$, which has moved from $1.8$ using the original PAO data to $5$ using the latest data. As a result, the KT diffuse neutrino flux has only changed marginally between the old and new PAO data set. To obtain the diffuse flux, the source distribution is integrated up to $z = 5$. This relation between the diffuse neutrino flux and the flux from Cen A is the main result of the KT model.

In their paper \cite{Koers:2008hv}, Koers \& Tinyakov used a model by Cuoco \& Hannestad \cite{:2007qd} to describe the neutrino emission from Cen A, $\phi_\nu^{\text{Cen A}}$, itself based on a model by Mannheim, Protheroe \& Rachen \cite{Mannheim:1998wp}. In this model, it is assumed that high-energy protons, accelerated by some mechanism (e.g., shock acceleration) are confined within a region close to the source. Because of energy losses in their photopion interactions with the ambient photon field, which is assumed to have an energy spectrum $n\left(E_\gamma\right) \varpropto E_\gamma^{-2}$, their lifetime is much shorter than their diffusive escape time and they decay into neutrons and neutrinos, both of which escape the source. Thereafter, the neutrons decay into UHECR protons; however, because of their interaction with the photon field before decaying, the neutrons produce a softer proton spectrum than the seed proton spectrum. Furthermore, the model predicts two spectral breaks in the CR spectrum, at energies at which the optical depths for proton and neutrino photopion production become unity. These two breaks are close in energy, though, so that, to simplify the model, only one spectral break is considered, at energy $E_\text{br}$. Below $E_\text{br}$, the UHECR proton and neutrino spectra are harder than the seed proton spectrum by one power of the energy, while above $E_\text{br}$, the UHECR proton spectrum is softer than the seed spectrum by one power of the energy and the neutrino spectrum is harder by one power of the energy. Hence, at high energies, the model predicts a neutrino spectrum that is harder by one power of the energy than the UHECR proton spectrum.

Following \cite{Koers:2008hv,:2007qd,Mannheim:1998wp}, the all-flavour neutrino spectrum from Cen A can be written as
\begin{equation}\label{EqFluxKTCenA1}
 \phi_{\nu_\text{all}}^{\text{Cen A}}\left(E_\nu\right) 
 = \frac{\xi_\nu}{\xi_n \eta_{\nu n}^2} \min\left( \frac{E_\nu}{\eta_{\nu n} E_{\text{br}}} , \frac{E_\nu^2}{\eta_{\nu n}^2 E_{\text{br}}^2} \right)
   \phi_p^{\text{Cen A}} \left( \frac{E_\nu}{\eta_{\nu n}} \right) ~,
\end{equation}
where $\xi_i$ ($i = \nu, n$) is the fraction of the proton's energy that is transferred to the species $i$ in photopion interactions and $\eta_{\nu n}$ is the ratio of the average neutrino energy to the average neutron energy. The KT model uses for these parameters the values featured in \cite{Mannheim:1998wp}, obtained from Monte Carlo simulations: $\xi_\nu \approx 0.1$, $\xi_n \approx 0.5$, $\langle E_\nu \rangle/E_p \approx 0.033$ and $\langle E_n \rangle / E_p \approx 0.83$, with which $\xi_\nu / \xi_n = 0.2$ and $\eta_{\nu n} = 0.04$. The neutrino break energy, $E_{\text{br}}$, is estimated from the gamma-ray break energy as $E_{\text{br}} \simeq 3 \times 10^8 E_{\gamma,\text{br}}$. Ref.~\cite{Mannheim:1998wp} uses $E_{\gamma,\text{br}} = 200$ MeV, so that $E_\text{br} = 10^8$ GeV. Under the assumption of equal flavour ratios at Earth, eq.~(\ref{EqRatiosDet}), the $\nu_\mu + \overline{\nu}_\mu$ flux is $1/3$ the flux in eq.~(\ref{EqFluxKTCenA1}). Plugging the power-law proton spectrum, eq.~(\ref{EqProtonSpectrum}), with the normalisation constant for a point source, eq.~(\ref{EqNormConstPt}), into the eq.~(\ref{EqFluxKTCenA1}) yields
\begin{equation}
 \phi_{\nu_\mu}^{\text{Cen A}} \left(E_\nu\right)
 = \frac{\Phi_p^{\text{Cen A}}\left(E_{\text{th}}\right)}{3} \frac{\xi_\nu \eta_{\nu n}^{\alpha_p-2}}{\xi_n} \frac{\alpha_p-1}{E_{\text{th}}}
   \left( \frac{E_\nu}{E_{\text{th}}} \right)^{-\alpha_p} \left( \frac{E_\nu}{E_{\nu,\text{br}}} \right) \min\left(1,\frac{E_\nu}{E_{\nu,\text{br}}}\right)
\end{equation}
for the muon-neutrino flux from Cen A, with $E_{\nu,\text{br}} \equiv \eta_{\nu n} E_{\text{br}} = 4 \times 10^6$ GeV. Using the scaling relation, eq.~(\ref{EqScalingKT}), the muon-neutrino diffuse flux in the KT model is therefore $\phi_{\nu_\mu}^\text{diff,KT} \left(E_\nu\right) \simeq 5 H \left(E_\text{th}\right) \phi_{\nu_\mu}^{\text{Cen A}} \left( E_\nu \right)$ and we can write it as
\begin{equation}\label{EqFluxDiffKT}
 \phi_{\nu_\mu}^\text{diff,KT} \left(E_\nu\right) 
 = A_\nu^\text{diff,KT} E_\nu^{-\alpha} \min\left(\frac{E_\nu}{E_{\nu,\text{br}}},\frac{E_\nu^2}{E_{\nu,\text{br}}^2}\right) ~,
\end{equation}
with the neutrino normalisation given by
\begin{equation}\label{EqAnuDiffKT}
 A_\nu^\text{diff,KT} \simeq \frac{5}{3} H\left(E_\text{th}\right) \frac{\xi_\nu}{\xi_n} \eta_{\nu n}^{\alpha_p-2}
 A_p^\text{Cen A}\left(\alpha_p\right) ~,
\end{equation}
and, following eq.~(\ref{EqNormConstPt}), $A_p^\text{Cen A} = \Phi_p^\text{Cen A} \left(\alpha_p-1\right) E_\text{th}^{\alpha_p-1}$.

\subsection{Model by Becker \& Biermann}\label{Section_Theory_Sub_BB}

The BB model \cite{Becker:2008nf} describes the production of high-energy neutrinos in the relativistic jets of radio galaxies. According to the model, the UHECRs observed by the PAO originated at FR-I galaxies (relatively low-luminosity radio galaxies with extended radio jets, and radio knots distributed along them), which can in principle accelerate protons up to about $10^{20}$ eV. Like in the KT model, here the protons are also shock-accelerated. Unlike the KT model, though, where the neutrino emission occurred in a region close to the AGN core, in the BB model the neutrino emission from $p\gamma$ interactions is expected to peak at the first strong shock along the jet, lying at a distance of $z_j \sim 3000$ gravitational radii from the center. 

The optical depth corresponding to proton interactions with the disc photon field $\tau_{p \gamma_{disc}} \approx 0.02$ and so $p\gamma$ interactions in the disc are not the dominant source of neutrinos. The proton-proton interactions that occur when the jet encounters the AGN's torus are also neglected as neutrino source in the BB model. The dominant mechanism of neutrino production is the interaction between the accelerated protons and the synchrotron photons in the relativistic jet, at one of the jet's knots. For boost factors of the streaming plasma of $\Gamma \sim 10$, the optical depth $\tau_{p \gamma_\text{synch}} \sim 1$. 

Hence, it is expected that neutrino emission occurs predominantly at the foot of the jet, where the beam is still highly collimated. Therefore, the BB model predicts a highly beamed neutrino emission, produced in the first shock ($z_j \sim 3000 r_g$), and consequently observable only from sources whose jets are directed towards Earth. Flat-spectrum radio sources, such as FR-I galaxies whose jets are pointing towards Earth, will have correlated neutrino and proton spectra, while steep-spectrum sources, which are AGN seen from the side, are expected to be weak neutrino sources, but to contribute to the cosmic-ray proton flux.

The BB model assumes that the $N_\text{corr} = 29$ events that were observed by the PAO to have a positional correlation to sources in the VCV catalogue were indeed originated at AGN lying in the supergalactic plane. In order to relate the proton and neutrino normalisation constants, $A_p^\text{diff}$ and $A_\nu^\text{diff}$, we will use the connection between the proton and neutrino energy fluxes \cite{Becker:2008nf}, i.e.,
\begin{equation}
 j_\nu = \frac{\tau_{p\gamma}}{12} \frac{\Gamma_\nu}{\Gamma_\text{CR}} \frac{\Omega_p}{\Omega_\nu} \frac{n_\nu}{n_p}\left(z_\text{CR}^{\max}\right) j_p ~,
\label{flux_rel}
\end{equation}
where $\Omega_\nu$, $\Omega_\text{CR}$ are the solid angles of emission of neutrinos and cosmic rays, respectively, and $\Gamma_\nu$, $\Gamma_\text{CR}$ are the boost factors of neutrinos and cosmic rays, respectively. The parameter $z_\text{CR}^{\max}$ is the redshift of the farthest AGN that contribute to the cosmic-ray flux. The total number of neutrino (proton) sources, $n_\nu$ ($n_p$), is calculated by integrating the luminosity function of Willott \cite{Willott:2000dh} (Dunlop \& Peacock \cite{Dunlop:1990kf}) from $z_\text{CR}^{\min} = 0.018$ ($0.0008$) up to $z_\text{CR}^{\max}$. 

On the other hand, assuming a power-law behaviour for the diffuse differential flux of protons, eq.~(\ref{EqProtonSpectrum}), the energy flux results in 
\begin{equation}\label{EqFluxProton}
 j_p 
 = A_p \int_{E_{p,\min}}^{E_{p,\max}} E_p \frac{dN_p}{dE_p} ~dE_p 
 =
 \left\{\begin{array}{ll}
  A_p \left(\alpha_p-2\right)^{-1} E_{p,\min}^{-\alpha_p+2} & , ~\text{if}~ \alpha_p \ne 2 \\
  A_p \ln\left(E_{p,\max}/E_{p,\min}\right)                 & , ~\text{if}~ \alpha_p = 2
 \end{array}\right. ~,
\end{equation}
where the term proportional to $E_{p,\max}^{-\alpha_p+2}$ has been neglected, in the case when $\alpha_p \ne 2$. Assuming that the neutrino spectrum follows the proton spectrum, i.e., $\phi_{\nu_\mu}^\text{diff,BB} = A_\nu^\text{diff} E_\nu^{-\alpha_\nu}$ with $\alpha_\nu \approx \alpha_p$, the energy flux for neutrinos is
\begin{equation}\label{EqFluxNeutrino}
 j_\nu \simeq
 \left\{\begin{array}{ll}
  A_\nu^\text{diff} \left(\alpha_p-2\right)^{-1} E_{\nu,\min}^{-\alpha_p+2} & , ~\text{if}~ \alpha_p \ne 2 \\
  A_\nu^\text{diff} \ln\left(E_{\nu,\max}/E_{\nu,\min}\right)               & , ~\text{if}~ \alpha_p = 2
 \end{array}\right. ~.
\end{equation}
The lower integration limits for protons and neutrinos are, respectively, $E_{p,\min} = \Gamma_p m_p \approx \Gamma_p \cdot \left(1~\text{GeV}\right)$ and $E_{\nu,\min} = \Gamma_\nu \cdot \left(m_\pi/4\right) = \Gamma_\nu \cdot (0.035~\text{GeV})$. Finally, 
replacing eq.~(\ref{EqFluxProton}), eq.~(\ref{EqFluxNeutrino}), and the proton normalisation constant $A_p^\text{diff}$ given by eq.~(\ref{EqNormConstDiff}) evaluated with $N_\text{evts} = N_\text{corr}$, we see that when $\alpha_p \ne 2$, the neutrino normalisation constant is 
\begin{equation}\label{EqAnu}
 A_\nu^\text{diff,BB} \simeq \frac{\tau_{p\gamma}}{12} \left(\frac{\Gamma_\nu}{\Gamma_p}\right)^{\alpha_p+1} \frac{n_\nu}{n_p}\left(z_\text{CR}^{\max}\right) \left(\frac{m_\pi}
              {4}\right)^{\alpha_p-2} A_p^\text{diff}\left(\alpha_p\right) ~.
\end{equation}
The dependence of $n_\nu/n_p$ on $z_\text{CR}^{\max}$ is shown graphically in Figure 5 of Ref.~\cite{Becker:2008nf}: $n_\nu/n_p$ decreases with $z_\text{CR}^{\max}$. To arrive at this expression\footnote{The reader should be wary that in their paper \cite{Becker:2008nf}, Becker \& Biermann incorrectly reported a dependence of the form $\sim \left(\Gamma_\nu/\Gamma_\text{CR}\right)^{5-\alpha_p}$ due to an algebraic mistake\cite{Becker:2009}.}, it must be noted that because of the relativistic beaming in the jets, the emission solid angles are $\Omega_\nu \sim 1/\Gamma_\nu^2$ and $\Omega_p \sim 1/\Gamma_\text{CR}^2$. When $\alpha_p = 2$, the logarithms in the two spectra are similar and cancel out, making the previous expression for $A_\nu^\text{diff,BB}$ valid also for $\alpha_p = 2$. Note that, since the ratio $N_\text{corr}/\Xi$ has decreased approximately by a factor of $2$ between the original and updated PAO analyses, then the updated BB diffuse flux is about half the original.

\section{Current and preliminary bounds on the neutrino flux}\label{Section_Bounds}

In the present work, we have assumed that the UHE AGN neutrino flux accounts for all of the UHE neutrino flux. This is, of course, a simplifying assumption, since high-energy contributions could also originate at other types of sources, such as gamma-ray bursts \cite{Waxman:1997ti,Bahcall:2000sa,Meszaros:2006vb,Abbasi:2009kq}.

We have taken into account three experimental bounds on the diffuse astrophysical neutrino flux: two upper bounds, one set by the AMANDA-II experiment and the other by its successor, IceCube, in its half-completed configuration of 40 DOM strings; and a lower bound given by the discovery potential of the final 86-string IceCube configuration. These three bounds have been included in figure \ref{FigFluxesFilled}. 

The AMANDA-II upper bound on the diffuse high-energy flux of extra-terrestrial muon-neutrinos was obtained by using data recorded between the years 2000 and 2004 \cite{Achterberg:2007qp}:
\begin{equation}\label{EqAMANDAexcl}
 E_\nu^2 \phi_{\nu_\mu}^\text{diff} \leq 7.4 \times 10^{-8} ~\text{GeV} ~\text{cm}^{-2} ~\text{s}^{-1} ~\text{sr}^{-1} \left(90\% ~\text{C.L.}\right)~,
\end{equation}
in the range $16$ TeV \--- $2.5$ PeV. This bound was set using exclusively upgoing UHE neutrinos, six of which were detected during the 807 days of live time reported. 

More recently, the IceCube Collaboration presented a preliminary upper bound using 375 days of recorded upgoing data with the half-completed IceCube-40 array which is almost an order of magnitude tighter than the AMANDA bound \cite{Grullon:2010fa}:
\begin{equation}\label{EqIC40excl}
 E_\nu^2 \phi_{\nu_\mu}^\text{diff} \leq 8 \times 10^{-9} ~\text{GeV} ~\text{cm}^{-2} ~\text{s}^{-1} ~\text{sr}^{-1} \left(90\% ~\text{C.L.}\right)~,
\end{equation}
in the range $10^{4.5}$ \--- $10^7$ GeV. 

Finally, the discovery potential at the $5\sigma$ level of the full, 86-string, IceCube array has been recently estimated \cite{Grullon:2010} to reach, after five years of exposure, 
\begin{equation}\label{EqIC86disc}
 E_\nu^2 \phi_{\nu_\mu}^\text{diff} \leq 7 \times 10^{-9} ~\text{GeV} ~\text{cm}^{-2} ~\text{s}^{-1} ~\text{sr}^{-1} \left(5\sigma\right)~,
\end{equation}
also in the range $10^{4.5}$ \--- $10^7$ GeV. This is the estimated minimum necessary flux required for a $5\sigma$ discovery after five years of running IC86. We will use this discovery potential as a lower bound on the neutrino flux. The discovery potential in eq.~(\ref{EqIC86disc}) is better than the original estimate of $9.9 \times 10^{-9} ~\text{GeV} ~\text{cm}^{-2} ~\text{s}^{-1} ~\text{sr}^{-1}$ that was presented in \cite{Ahrens:2003ix} due to a better knowledge of the detector and improved simulations.

Note that these three bounds were obtained under the assumption of an $E_\nu^{-\alpha}$ neutrino flux, with $\alpha=2$. For the KT and BB models in our work, however, we have allowed for $\alpha \neq 2$. Therefore, we have calculated for each one of them the associated number of muon-neutrinos in the AMANDA, IceCube-40 and IceCube-86 configurations, as appropriate, by assuming an $E_\nu^{-2}$ flux, and used these derived bounds on the number of events, and not on the flux, to constrain the KT and BB models. Concretely, we have assumed a $\phi_{\nu_\mu}^\text{diff}\left(E_\nu\right) = k E_\nu^{-2}$ flux, with the normalisation, $k$, given in each case 
by the numerical value of the bounds in eqs.~(\ref{EqAMANDAexcl})\---(\ref{EqIC86disc}), 
in units of GeV$^{-1}$ cm$^{-2}$ s$^{-1}$ sr$^{-1}$. These numbers are displayed in table \ref{tblNumEvtsBounds}. The expressions required to calculate the number of upgoing muon-neutrinos in the AMANDA, IceCube-40, and IceCube-86 arrays, for an arbitrary diffuse neutrino flux $\phi_{\nu_\mu}^\text{diff}$, are contained in Appendix \ref{Section_App_IceCube}. We have assumed that the effective detector area of AMANDA is $1/100$ times that of IceCube-86 and that the effective area of IceCube-40 is half the area of IceCube-86, on account of half the number of strings having been deployed. Note, however, that this is only an estimate, since the actual effective area of IceCube-40 will be strongly dependent on the efficiency of the cuts employed to calculate it.

\section{Muon-neutrino number of events in the IceCube-86 detector for the BB and KT models}\label{Section_NumEvts}

\begin{table}[t!]
 \begin{center}
  \small
  \begin{tabular}{|l|c|c|c|c|}
  \hline
  Limit                                                               & Energy range [GeV]              & Exp. time & Upgoing $\nu_\mu$ \\
  \hline
  AMANDA upper bound (AMANDA) \cite{Achterberg:2007qp}                & $1.6\times10^4\--2.5\times10^6$ & 807 days  & 6.0    \\
  IceCube-40 preliminary upper bound (IC40) \cite{Grullon:2010fa}     & $10^{4.5}\--10^7$               & 375 days  & 5.90   \\
  IceCube-86 estimated $5\sigma$ discovery (IC86) \cite{Grullon:2010} & $10^{4.5}\--10^7$               & 5 years   & 50.28  \\ 
  \hline
  \end{tabular}
  \caption{Maximum number of upgoing muon-neutrinos allowed by the reported exclusion limit from AMANDA and the preliminary one from IceCube-40, and minimum number of events needed for $5\sigma$ discovery according to the estimated IceCube-86 5-year discovery flux. In every case, the event numbers were calculated by assuming a $E_\nu^{-2}$ diffuse flux. Each bound on the number of events was calculated in the respective detector configuration (Appendix \ref{Section_App_IceCube} contains the effective area for each), with the corresponding exposure time.\label{tblNumEvtsBounds}}
 \end{center}
\end{table}

\begin{table}[t!]
 \begin{center}
  \begin{tabular}{|c|c|c|}
   \hline
   Limit  & no source evolution & strong source evolution \\ 
   \hline
   AMANDA & 3.04                & 2.81                    \\
   IC40   & 2.59                & 2.27                    \\ 
   IC86   & 2.57                & 2.25                    \\
   \hline
  \end{tabular}
  \caption{Maximum value of the spectral index $\alpha$ in the Koers-Tinyakov model allowed by the upper bounds AMANDA and IC40, and minimum value needed for $5\sigma$ discovery according to the estimated IC86 discovery potential.\label{tblAlphaBounds}}
 \end{center}
\end{table}

In this Section, we study the Koers \& Tinyakov (KT) and the Becker \& Biermann (BB) models of diffuse AGN neutrino flux through their predictions of the number of muon-neutrinos that will be detected by the full IceCube-86 neutrino detector. To calculate the number of neutrinos, we have adopted the method followed in \cite{Koers:2008hv}, which is summarised here in Appendix \ref{Section_App_IceCube}. 

In our analysis, we have fixed the IceCube-86 detector exposure time at $T = 5$ years and calculated the integrated event yield within the energy range $10^5\--10^8$ GeV. Only upgoing neutrinos have been considered, i.e., those that reach the detector with zenith angles between $90^\circ$ and $180^\circ$ (the normal to the South Pole lies at $0^\circ$), for which the atmospheric neutrino and muon background is filtered out by interactions inside the Earth. Downgoing neutrinos, i.e., those with zenith angles between $0^\circ$ and $90^\circ$, traverse only about $10$ km of atmosphere before reaching the detector and have not been included in the analysis due to the added difficulty of separating the atmospheric background from the astrophysical neutrino signal. Furthermore, in the case of the KT flux, we have considered both the scenario with no source evolution and the one with strong source evolution. 

Based on the experimental bounds introduced in the previous Section, we have defined two visibility criteria with the purpose of identifying the regions of parameter space allowed by the upper limits and accessible by the discovery potential of the full IceCube-86 array. Under the first one \---the AMANDA visibility criterion\---, the IceCube-86 event-rate predictions, for either KT or BB, are required to lie above the IC86 discovery potential and below the AMANDA upper bound. Similarly, under the IC40 visibility criterion, the event rates must lie above the IC86 discovery potential and below the IC40 upper bound. 

\subsection{Parameters under study and neutrino fluxes}

We have calculated our expectations of the neutrino flux models taking as free parameters $\alpha$ for the KT model (to simplify, we will use $\alpha \equiv \alpha_p$ hereafter), and $\alpha$, $\Gamma_\nu/\Gamma_\text{CR}$, and $z_\text{CR}^{\max}$ for the BB model, and varied them within the following intervals:
\begin{equation}
 2 \le \alpha \le 3 \ , \qquad
 1 \le \Gamma_\nu/\Gamma_\text{CR} \le 20 \ , \qquad
 10^{-3} \le z_\text{CR}^{\max} \le 0.03 ~.
\end{equation}
This range of $\alpha$ has been chosen in order to cover a wide range around 2.7, the preferred value obtained from fits to combined cosmic-ray data \cite{Berezinsky:2002nc}, or values less than 2.3 that are predicted in case of stochastic shock acceleration \cite{Baring:2004qc, Bednarz:1998pi,Kardashev:1962,Meli:2007zc,Meli:2007sv}. We have defined the range of $\Gamma_\nu/\Gamma_\text{CR}$ for values greater than 1 since, under the assumptions made by the BB model, the neutrinos are produced in early shocks and protons, in late ones. Besides, it includes the value of $3$ used in \cite{Becker:2008nf}. The range of $z_\text{CR}^{\max}$ is the same as the one used in said reference. Our purpose in varying the latter parameter, $z_\text{CR}^{\max}$, is to test different hypotheses about the maximum redshift up to which the AGN contribute to the UHE diffuse neutrino flux. We remind the reader that the results for the KT model have been obtained for a fixed value of $z_\text{CR}^{\max}=5$ and so they were not affected by this variation.  

\begin{figure}[t!]
 \begin{center}
  \scalebox{0.7}{\includegraphics{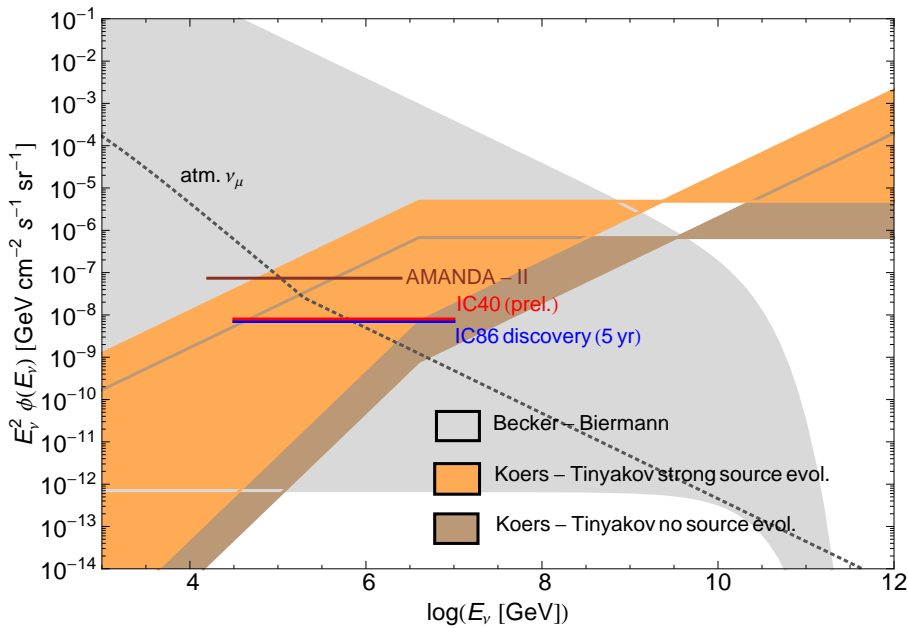}} 
  \caption{AGN muon-neutrino fluxes, multiplied by $E_\nu^2$, according to the models by Becker \& Biermann (BB) and Koers \& Tinyakov (KT), with strong source evolution and without source evolution. The regions were generated by varying the model parameters in the ranges $2 \le \alpha \le 3$, $1 \le \Gamma_\nu/\Gamma_\text{CR} \le 20$, and $10^{-3} \le z_\text{CR}^{\max} \le 0.03$. The grey region corresponds to all the possible BB fluxes resulting from the variation of $\alpha$, $\Gamma_\nu/\Gamma_\text{CR}$, and $z_\text{CR}^{\max}$, whereas the brown and orange regions correspond to all the possible KT fluxes resulting from the variation of $\alpha$, under the assumption of no source evolution and of strong source evolution, respectively. The atmospheric muon-neutrino flux has been plotted (in black, dotted, lines) for comparison. The AMANDA-II upper bound, the preliminary 40-string IceCube upper bound and an estimated 86-string IceCube five-year discovery potential at $5\sigma$ have been included by assuming a $E_\nu^{-2}$ flux. The atmospheric neutrino flux is given by the parametrisation in Ref.~\cite{Koers:2008hv}. See the text for details.}
  \label{FigFluxesFilled}
 \end{center}
\end{figure}

Figure \ref{FigFluxesFilled} shows the BB and KT diffuse muon-neutrino fluxes, multiplied by $E_\nu^2$, as functions of the neutrino energy, when the values of the model parameters are varied within the ranges that we have quoted above. We have also included the upper bounds on the flux set by AMANDA and IceCube-40, and the estimated discovery potential of IceCube-86 after five years of running. Our analysis will focus on the different regions enclosed between these upper bounds and the IC86 discovery potential taken as a lower bound, in the energy range $10^5$\---$10^8$ GeV, where the fluxes may be detected in IceCube. We will find how the bounds on the neutrino flux translate into bounds on the values of $\alpha$, $\Gamma_\nu/\Gamma_\text{CR}$, and $z_\text{CR}^{\max}$, thus restricting the capacity of the KT and BB flux models to account for an observed extra-terrestrial neutrino signal.

\subsection{KT event-rate expectations in IceCube-86} 

Since the KT flux depends on a single parameter, i.e., the spectral index $\alpha$, we can translate the bounds on event numbers directly into bounds on $\alpha$. In this way, the results presented in table \ref{tblAlphaBounds} represent the upper limits on $\alpha$ given by the AMANDA and IC40 bounds and the lower limits given by the IC86 discovery potential.

\begin{figure}[t!]
 \begin{center}
  \scalebox{0.6}{\includegraphics{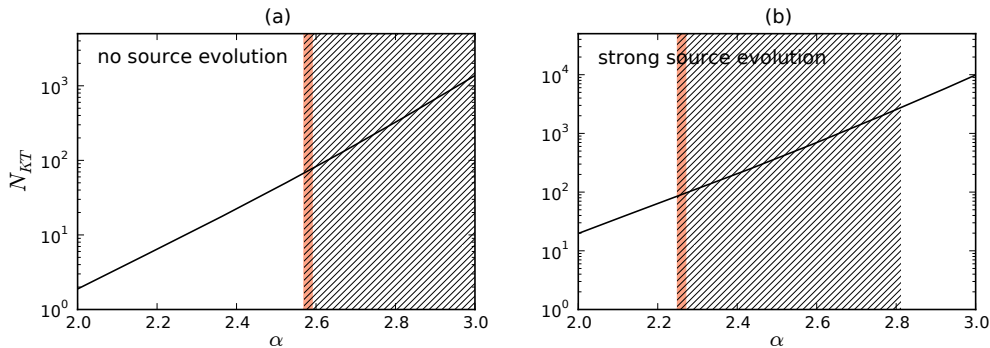}}
  \caption{Integrated number of upgoing muon-neutrinos, between $10^5$ and $10^8$ GeV, expected in IceCube-86, after $T = 5$ years of exposure, associated to the KT production model assuming (a) no source evolution and (b) strong source evolution. The orange-coloured bands are the regions of values of $\alpha$ that lies above the IC86 discovery potential and below the IC40 upper bound, while the hatched region lies above IC86 and below the AMANDA upper bound (see table \ref{tblAlphaBounds}).}
  \label{Fig.KT.nevts.vs.alpha.array}
 \end{center}
\end{figure}

Figure \ref{Fig.KT.nevts.vs.alpha.array} shows the integrated number of upgoing muon-neutrinos with energies between $10^5$ and $10^8$ GeV, as a function of $\alpha$, that is expected in the full 86-string IceCube array after five years of exposure. Plot (a) assumes no source evolution, whereas (b) assumes strong source evolution. The predictions under the assumption of strong source evolution are up to an order of magnitude higher than under no source evolution. This fact can be easily understood since a difference of a similar magnitude is found in the neutrino boost factor, as shown in Ref.~\cite{Koers:2008hv}.

The orange-coloured and hatched bands mark the visibility regions under the IC40 and AMANDA visibility criteria, respectively, according to table \ref{tblAlphaBounds}. Owing to the fact that the AMANDA upper bound is less restrictive than the IC40 bound, the visibility regions are in every case larger when the former one is used. According to figure \ref{Fig.KT.nevts.vs.alpha.array} and table \ref{tblAlphaBounds}, the ranges of event numbers, $N_\text{KT}$, that IceCube-86 will be able to detect in the interval $10^5\--10^8$ GeV, after five years of exposure, are:
\begin{equation}\label{EqNKTVisRegionNoEvol}
 68 \le N_\text{KT}^\text{up} \le 77 \left(1847\right) ~,
\end{equation}
assuming no source evolution and using the IC40 (AMANDA) upper bound, and
\begin{equation}\label{EqNKTVisRegionEvol}
 85 \le N_\text{KT}^\text{up} \le 95 \left(2709\right) ~,
\end{equation}
assuming strong source evolution.

From table \ref{tblAlphaBounds}, we see that the KT model with no source evolution is allowed for higher values of $\alpha$ than the model with strong source evolution. This is due to the fact that the KT flux grows with $\alpha$, and that, for a given value of $\alpha$, the event yield produced by the strong source evolution model is up to an order of magnitude higher than the yield with no source evolution. Thus, lower values of $\alpha$ are needed to keep the former below the IC40 or AMANDA event-number upper bounds. 

From the same table, we find that for the KT model the value of $\alpha = 2.7$, obtained from fits to cosmic-ray data, would still be allowed under the AMANDA visibility criterion, but is discarded by the more recent IC40 criterion, regardless of the choice of source evolution. Under the assumption of strong source evolution, the other proposed value of $\alpha = 2.3$ is excluded (permitted) by the IC40 (AMANDA) visibility criterion, while values of $\alpha \leq 2.25$ would be out of reach of the IceCube discovery potential. Under no source evolution, the region of $\alpha$ below the IC86 potential starts from $2.57$. This constitutes a strong hint toward the KT flux being too large. However, as explained in Section \ref{Section_Bounds}, we would like to stress that our visibility criteria make use of event-yield bounds that are deduced from bounds on a $E_\nu^{-2}$ flux, a comparison that might be overly reducing the size of the visibility regions. A more sophisticated analysis that makes use of model-independent flux bounds, i.e., bounds not exclusive to $E_\nu^{-2}$ models, will be presented elsewhere \cite{Arguelles:2011}.

\subsection{BB event-rate expectations in IceCube-86} 

\begin{figure}[t!]
 \begin{center}
  \scalebox{0.6}{\includegraphics{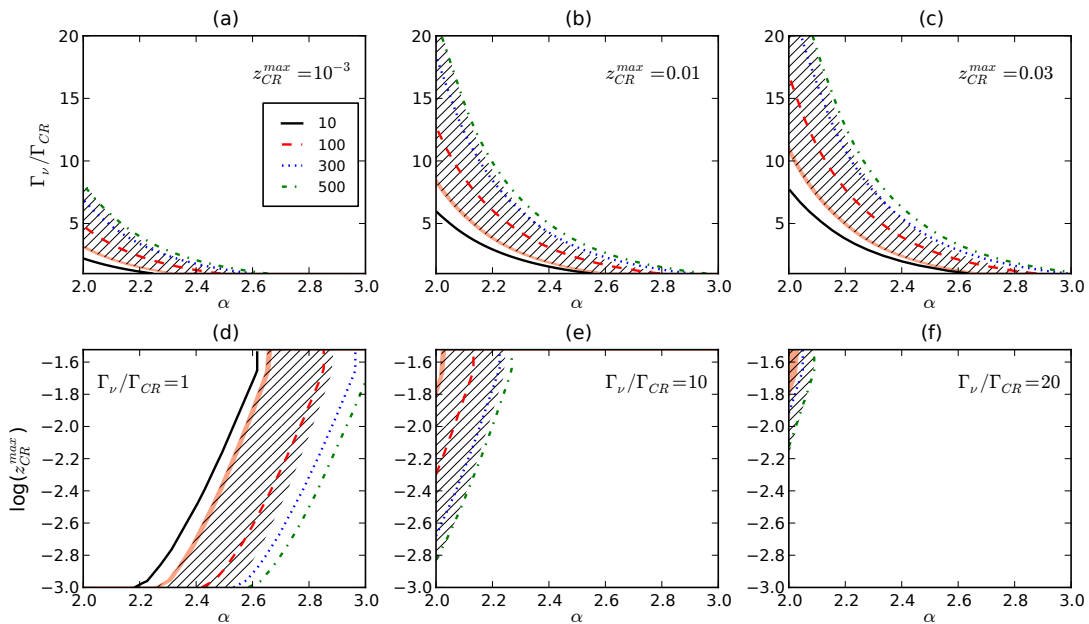}}
  \caption{Variation of the integrated number of upgoing muon-neutrinos expected in the range $10^5 \leq E_\nu/\text{GeV} \leq 10^8$ associated to the BB model, after $T = 5$ years of exposure of the IceCube-86 detector. In (a), (b), and (c), the value of $z_\text{CR}^{\max}$ has been fixed, respectively, at the representative values of $10^{-3}$, $0.01$, and $0.03$, while $\alpha$ and $\Gamma_\nu/\Gamma_\text{CR}$ have been allowed to vary. Likewise, in (d), (e), and (f), $\Gamma_\nu/\Gamma_\text{CR}$ has been fixed at $1$, $10$, and $20$, respectively, while $\alpha$ and $z_\text{CR}^{\max}$ have been varied. The solid lines are iso-contours of number of events: $10$ (solid black), $100$ (dashed red), $300$ (dotted blue), and $500$ (dash-dotted green). The region coloured orange is the parameter region where the event-number predictions lie above the IC86 discovery potential and below the IC40 upper bound, i.e., the IC40 visibility region. Similarly, the hatched region is where the predictions lie above the IC86 potential and below the AMANDA upper bound, i.e., the AMANDA visibility region.}
  \label{Fig.nevts.Emin.1.D5.Emax.1.D8.UP.array}
 \end{center}
\end{figure}

As to the BB flux model, figure \ref{Fig.nevts.Emin.1.D5.Emax.1.D8.UP.array} shows iso-contours of the expected integrated number of upgoing muon-neutrinos in the IceCube-86 detector, in the $\Gamma_\nu/\Gamma_\text{CR}$\---$\alpha$ plane, for fixed values of (a) $z_\text{CR}^{\max} = 10^{-3}$, (b) $0.01$, and (c) $0.03$, and in the $z_\text{CR}^{\max}$\---$\alpha$ plane, for fixed values of (d) $\Gamma_\nu/\Gamma_\text{CR} = 1$, (e) $10$, and (f) $20$. The BB normalisation constant, according to eq.~(\ref{EqAnu}), decreases with $z_\text{CR}^{\max}$ and increases with $\Gamma_\nu/\Gamma_\text{CR}$. This behaviour is observed in figure \ref{Fig.nevts.Emin.1.D5.Emax.1.D8.UP.array}, where, for fixed values of $\alpha$ and $\Gamma_\nu/\Gamma_\text{CR}$, the number of events decreases as $z_\text{CR}^{\max}$ increases. On the other hand, for fixed values of $\alpha$ and $z_\text{CR}^{\max}$, the number of events increases with $\Gamma_\nu/\Gamma_\text{CR}$.   
 
In each plot, as we have mentioned before, the IC40 visibility region is coloured orange and lies between the IC86 discovery potential (left border) and the IC40 upper bound (right border) listed in table \ref{tblNumEvtsBounds}. The AMANDA visibility region, on the other hand, is represented by the hatched region, and its right border is fixed instead by the AMANDA bound. 

Besides the observed narrowness of the visibility regions, there are two main features to point out. First, if the value of $z_\text{CR}^{\max}$ increases, the allowed ranges of $\alpha$ and $\Gamma_\nu/\Gamma_\text{CR}$ also increase, with higher values being allowed. Second, if the value of $\Gamma_\nu/\Gamma_\text{CR}$ increases, the allowed ranges of $\alpha$ and $z_\text{CR}^{\max}$ decrease, with $\alpha$ tending to lower values and $z_\text{CR}^{\max}$ to higher ones. These observations can be quantified if we project the visibility regions in each plane onto the horizontal and vertical axes. The allowed regions of the parameters are shown in tables \ref{tblBBBounds1} and \ref{tblBBBounds2}.

\begin{table}[t!]
 \begin{center}
  \small
  \begin{tabular}{|c|c|c|c|c|c|c|}
  \hline
  $z_\text{CR}^{\max}$  & Minimum  & \multicolumn{2}{|c|}{Maximum $\alpha$} & Minimum & \multicolumn{2}{|c|}{Maximum $\Gamma_\nu/\Gamma_\text{CR}$} \\
                         \cline{3-4} \cline{6-7}
                        & $\alpha$ & IC40 visib. & AMANDA visib.    & $\Gamma_\nu/\Gamma_\text{CR}$ & IC40 visib. & AMANDA visib. \\
  \hline
  $10^{-3}$ & $2$ & $2.3$ & $2.65$ & $1$ & $3$ & $7.5$     \\
  $0.01$    & $2$ & $2.6$ & $2.95$ & $1$ & $8.5$ & $20$ \\
  $0.03$    & $2$ & $2.65$ & $3$ & $1$ & $11$ & $20$ \\
  \hline
  \end{tabular}
  \caption{Allowed intervals of $\alpha$ and $\Gamma_\nu/\Gamma_\text{CR}$ obtained by projecting the visibility regions from plots \ref{Fig.nevts.Emin.1.D5.Emax.1.D8.UP.array}a\---c onto the axes.\label{tblBBBounds1}}
 \end{center}
\end{table}

\begin{table}[t!]
 \begin{center}
  \small
  \begin{tabular}{|c|c|c|c|c|c|c|}
  \hline
  $\Gamma_\nu/\Gamma_\text{CR}$  & Minimum  & \multicolumn{2}{|c|}{Maximum $\alpha$} & \multicolumn{2}{|c|}{Minimum $z_\text{CR}^{\max}$} & Maximum \\
                         \cline{3-4} \cline{5-6}
                        &   $\alpha$        & IC40 visib. & AMANDA visib.    & IC40 visib. & AMANDA visib. & $z_\text{CR}^{\max}$ \\
  \hline
  $1$   & $2.25$ & $2.65$ & $2.9$  & $10^{-3}$ & $10^{-3}$ & $0.03$     \\
  $10$  & $2$    & $2.03$ & $2.22$ & $0.015$ & $0.002$ & $0.03$ \\
  $20$  & $2$    & $2.03$ & $2.1$ & $0.015$ & $0.008$ & $0.03$ \\
  \hline
  \end{tabular}
  \caption{Allowed intervals of $\alpha$ and $z_\text{CR}^{\max}$ obtained by projecting the visibility regions from plots \ref{Fig.nevts.Emin.1.D5.Emax.1.D8.UP.array}d\---f onto the axes.\label{tblBBBounds2}}
 \end{center}
\end{table}

In light of the results presented in these tables, and momentarily assuming that $\alpha=2.7$ is the true value of the cosmic-ray spectral index \cite{Berezinsky:2002nc}, we see that under the AMANDA visibility criterion the BB flux model is clearly excluded for $\Gamma_\nu/\Gamma_\text{CR} \gtrsim 10$ (for any value of $z_\text{CR}^{\max}$) and also for the lowest values of $z_\text{CR}^{\max}$, close to $10^{-3}$ (for any value of $\Gamma_\nu/\Gamma_\text{CR}$). Whenever $\alpha=2.7$ is allowed by the BB model, it is only inside a very narrow region of parameter space, around $\Gamma_\nu/\Gamma_\text{CR} \sim 1$ and $z_\text{CR}^{\max} \gtrsim 0.004$. On the other hand, under the more recent IC40 visibility criterion, the BB model at $\alpha = 2.7$ is discarded for all values of $\Gamma_\nu/\Gamma_\text{CR}$ and $z_\text{CR}^{\max}$. 

If we consider the other values of $\alpha = 2.3$ and $2.0$ proposed in the literature (see 
Ref.~\cite{Becker:2008nf} and references therein), we find that the allowed regions, for $\alpha = 2.3$ and the AMANDA visibility criterion, are: $1 \lesssim \Gamma_\nu/\Gamma_\text{CR} \lesssim 3$, $2.5 \lesssim \Gamma_\nu/\Gamma_\text{CR} \lesssim 6.5$ and $3 \lesssim \Gamma_\nu/\Gamma_\text{CR} \lesssim 8$ for $z_\text{CR}^{\max} = 10^{-3}, 0.01$ and $0.03$, respectively. In the case of $\alpha = 2$, and the AMANDA visibility criterion, the allowed regions are: $3 \lesssim \Gamma_\nu/\Gamma_\text{CR} \lesssim 8$, $8 \lesssim \Gamma_\nu/\Gamma_\text{CR} \lesssim 20$ and $11 \lesssim \Gamma_\nu/\Gamma_\text{CR} \lesssim 20$ for $z_\text{CR}^{\max} = 10^{-3}, 0.01$ and $0.03$, respectively. For $\alpha = 2.3(2.0)$, and the IC40 visibility criterion, the allowed values for $\Gamma_\nu/\Gamma_\text{CR}$ are: $1(3),2.5(8)$ and $3(11)$ for $z_\text{CR}^{\max} = 10^{-3}, 0.01$ and $0.03$, respectively. Clearly, lower values of $\alpha$ fare better under the more recent IC40 upper bound. Like for the KT model, the BB model region of parameter space could be larger if an analysis based on non-$E_\nu^{-2}$ bounds were performed instead. 
                 
\section{Comparison between the KT and BB models using the IceCube detector}\label{Section_ModelCompare}

\begin{figure}[t!]
 \begin{center} 
  \scalebox{0.6}{\includegraphics{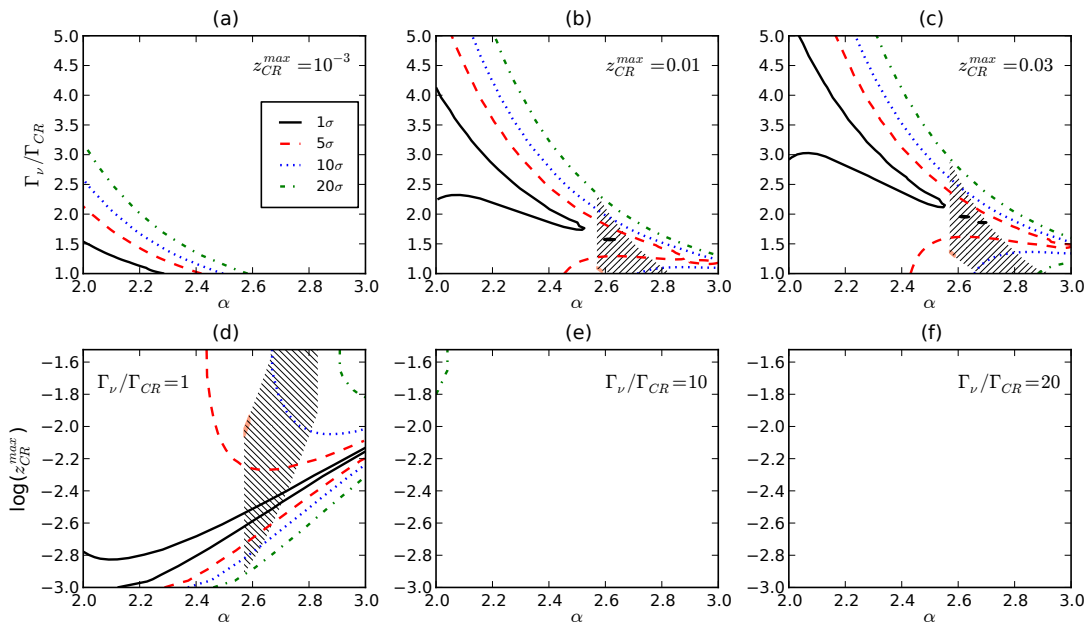}}
  \caption{Separation between the BB and KT models, in terms of $\Delta \equiv \left\vert N_\text{BB} - N_\text{KT}\right\vert$, measured in units of $\sigma \equiv \sqrt{N_\text{KT}}$ (see text), for upgoing neutrinos with energies in the range $10^5 \leq E_\nu/\text{GeV} \leq 10^8$ and assuming no source evolution for the KT model. The exposure time $T = 5$ years. In (a), (b), and (c), the value of $z_\text{CR}^{\max}$ has been fixed, respectively, at the representative values of $10^{-3}$, $0.01$, and $0.03$, while $\alpha$ and $\Gamma_\nu/\Gamma_\text{CR}$ have been allowed to vary. Likewise, in (d), (e), and (f), $\Gamma_\nu/\Gamma_\text{CR}$ has been fixed at $1$, $10$, and $20$, respectively, while $\alpha$ and $z_\text{CR}^{\max}$ have been varied. The solid lines are iso-contours of $\Delta = 1\sigma$ (solid black), $5\sigma$ (dashed red), $10\sigma$ (dotted blue), and $20\sigma$ (dash-dotted green). The region coloured orange is the parameter region where the event-number predictions of, simultaneously, the KT and BB models lie above the IC86 discovery potential and below the IC40 upper bound, i.e., the IC40 visibility region. Similarly, the hatched region is where the predictions of both models lie above the IC86 potential and below the AMANDA upper bound, i.e., the AMANDA visibility region. These regions of simultaneous visibility are where comparison between the two production models is meaningful, according to each of the two visibility criteria.}
  \label{Fig.sigmasep.Emin.1.D5.Emax.1.D8.noevol.UP.array}
 \end{center}
\end{figure}

\begin{figure}[t!]
 \begin{center}
 \scalebox{0.6}{\includegraphics{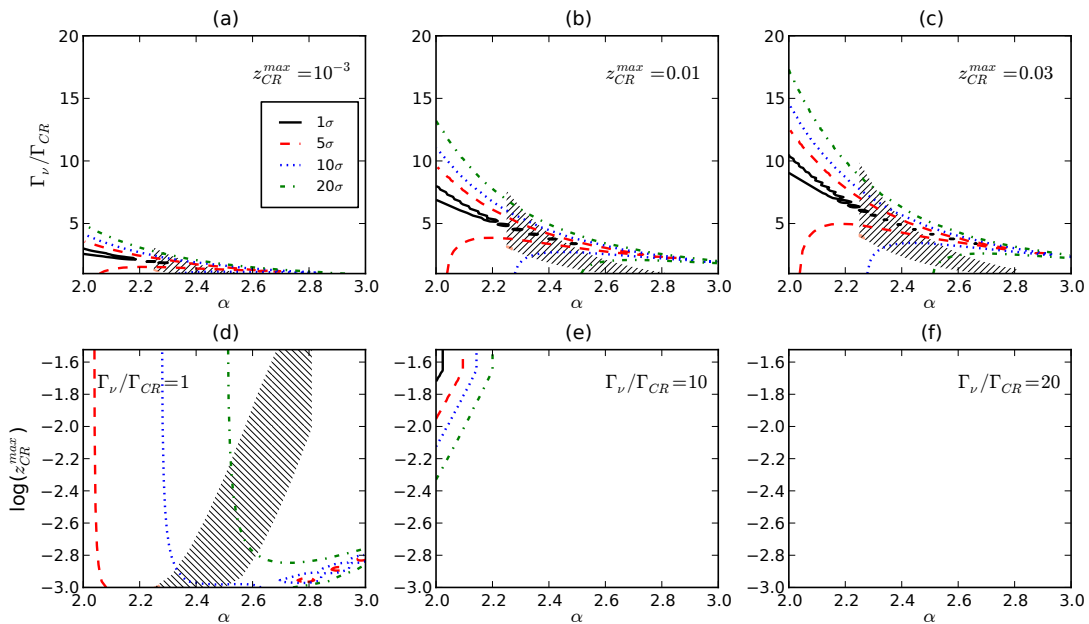}}
 \caption{Same as figure \ref{Fig.sigmasep.Emin.1.D5.Emax.1.D8.noevol.UP.array}, but assuming strong source evolution for the KT model.}
 \label{Fig.sigmasep.Emin.1.D5.Emax.1.D8.evol.UP.array}
 \end{center}
\end{figure}

We have quantified the difference between the predictions put forward by the two models using the quantity
\begin{equation}
 \Delta\left(\alpha,\Gamma_\nu/\Gamma_\text{CR},z_\text{CR}^{\max}\right)
 =
 \left\vert N_\text{BB}\left(\alpha,\Gamma_\nu/\Gamma_\text{CR},z_\text{CR}^{\max}\right) - N_\text{KT}\left(\alpha\right) \right\vert ~,
\end{equation}
and expressed it in units of $\sigma\left(\alpha\right) \equiv \sqrt{N_\text{KT}\left(\alpha\right)}$, i.e., at every point in parameter space we have measured the difference between the number of events predicted by each model, in units of the standard deviation of the KT prediction, assuming for it an uncertainty characteristic of a Gaussian distribution. The higher the value of $\Delta$, the greater the difference between the predictions. The comparison between the models, however, is only valid within the region that results from the intersection of the individual KT and BB visibility regions, given, respectively, by table \ref{tblAlphaBounds} and figure \ref{Fig.nevts.Emin.1.D5.Emax.1.D8.UP.array}. This guarantees that the numbers of events predicted by both models lie above the minimum required signal for detection at $5\sigma$ from the atmospheric neutrino background, so that the comparison between them is meaningful.

Figures \ref{Fig.sigmasep.Emin.1.D5.Emax.1.D8.noevol.UP.array} and \ref{Fig.sigmasep.Emin.1.D5.Emax.1.D8.evol.UP.array} show the separation between the models using the integrated number of muon-neutrinos in the IceCube-86 detector. The iso-contours correspond to $\Delta/\sigma = 1$ (solid black), $5$ (dashed red), $10$ (dotted blue), and $20$ (dash-dotted green), in the plane $\Gamma_\nu/\Gamma_\text{CR}$\---$\alpha$, for values of (a) $z_\text{CR}^{\max} = 10^{-3}$, (b) $0.01$, and (c) $0.03$, and in the plane $\log\left(z_\text{CR}^{\max}\right)$\---$\alpha$, for values of (d) $\Gamma_\nu/\Gamma_\text{CR} = 1$, (e) $10$, and (f) $20$. Where only one or none of the models are visible, the discrimination between them is obvious or meaningless, respectively. We have coloured orange the region of simultaneous visibility under the IC40 criterion, and hatched the region of simultaneous visibility under the AMANDA criterion. Evidently, since the individual visibility regions of the KT and BB models are larger under the AMANDA visibility criterion than under the IC40 criterion, the regions of simultaneous visibility are in every case larger under the former. 

We see that the KT and BB visibility regions overlap only at low values of $\Gamma_\nu/\Gamma_\text{CR}$ and that the size of the overlapping regions grows with $z_\text{CR}^{\max}$, so that they are largest for $z_\text{CR}^{\max} = 0.01$ and $0.03$, as shown in plots (b) and (c) of figures \ref{Fig.sigmasep.Emin.1.D5.Emax.1.D8.noevol.UP.array} and \ref{Fig.sigmasep.Emin.1.D5.Emax.1.D8.evol.UP.array}. In particular, under the AMANDA visibility criterion, and assuming no source evolution, the regions of simultaneous visibility exist only for low values of $\Gamma_\nu/\Gamma_\text{CR}$, between $1$ and $3$, while assuming strong source evolution, they exist up to $\Gamma_\nu/\Gamma_\text{CR} \approx 10$. Under the IC40 visibility criterion, comparison is allowed only inside very small regions of simultaneous visibility that lie at $\alpha \simeq 2.57 (2.25) \-- 2.59 (2.27)$, $\Gamma_\nu/\Gamma_\text{CR} \simeq 1 (3) \-- 1.5 (4)$, and $z_\text{CR}^{\max} = 0.01 \-- 0.03$, assuming no (strong) source evolution. Hence, comparison between the models becomes unfeasible in most of the parameter space. 

Regardless, within the small IC40 simultaneous visibility region, the models can be separated in no less than $5\sigma$ and no more than $10\sigma$, under both assumptions on source evolution, whereas under the dated AMANDA visibility criterion separations can vary between $1\sigma$ and $20\sigma$. Separations of $5\sigma$ would be sufficient to discern in a statistically meaningful way between the KT and BB models. Notice that the comparison at the favoured value of $\alpha = 2.7$ is not allowed under the IC40 visibility criterion, since neither flux will be visible in IceCube-86. For $\alpha = 2.0$ and $2.3$, there is no region of simultaneous visibility under this same visibility criterion.

\section{Summary and conclusions}\label{Section_Conclusions}

We have studied the IceCube-86 event rate expectations for two models of AGN diffuse muon-neutrino flux proposed in the literature, one by Koers \& Tinyakov (KT) \cite{Koers:2008hv} and another by Becker \& Biermann (BB) \cite{Becker:2008nf}, both of which take into account the apparent correlation, reported by the Pierre Auger Collaboration \cite{:2010zzj}, between the incoming directions of the highest-energy ($E > 55$ EeV) cosmic rays and the positions of AGN in the 12th edition V\'eron-Cetty \& V\'eron catalogue \cite{Veron:2006}. In doing this, we have assumed that the flux of neutrinos from AGN makes up all of the UHE astrophysical neutrino flux. Both models propose a power-law flux, i.e., proportional to $E_\nu^{-\alpha}$, resulting from shock acceleration.

In our analysis, we have taken the spectral index, $\alpha$, as well as two other parameters associated to the BB model, namely, the ratio of relativistic boost factors of neutrinos and cosmic rays, $\Gamma_\nu/\Gamma_\text{CR}$, and the redshift of the most distant AGN that contributes to the diffuse cosmic-ray flux, $z_\text{CR}^{\max}$, as free parameters, and varied their values within the following intervals: $2 \le \alpha \le 3$, $1 \le \Gamma_\nu/\Gamma_\text{CR} \le 20$, and $10^{-3} \le z_\text{CR}^{\max} \le 0.03$. In addition, we have explored the KT model under two assumptions on the evolution of the number density of AGN: either they do not evolve with redshift, or they evolve strongly with it, following the star formation rate. Neutrino fluxes calculated using the latter assumption are up to an order of magnitude higher than the ones calculated using the former one.

For each point $\left(\alpha,\Gamma_\nu/\Gamma_\text{CR},z_\text{CR}^{\max}\right)$ in parameter space, we have calculated for both models the associated integrated number of upgoing muon-neutrinos, between $10^5$ and $10^8$ GeV, that is expected after five years of exposure of the full 86-string IceCube neutrino detector (IceCube-86). In order to determine the regions of parameter space that this detector will be able to probe, we have tested two different upper bounds on the UHE neutrino flux: the bound reported by the AMANDA Collaboration using 807 days of observation \cite{Achterberg:2007qp} and a preliminary bound obtained after 375 days of exposure of the half-completed IceCube-40 detector (IC40) \cite{Grullon:2010fa}. A lower bound, on the other hand, was fixed at the estimated IceCube-86 five-year discovery potential at the $5\sigma$ level (IC86) \cite{Grullon:2010}. With this we have defined ``regions of visibility'' in parameter space as those regions inside which the event-rate predictions lie above the IC86 discovery potential and below the AMANDA or IC40 upper bound. Since the IC40 upper bound is lower than the AMANDA bound, the former restricts the allowed parameter space more than the latter. 

It is possible to confine the spectral index of the KT model within the range $2.57 \le \alpha \le 2.59 \left(3.04\right)$, under the assumption of no source evolution and using the IC40 (AMANDA) upper bound, and $2.25 \le \alpha \le 2.27 \left(2.81\right)$, under the assumption of strong source evolution. For the BB model, we found that IceCube-86 is sensitive to high values of $\Gamma_\nu/\Gamma_\text{CR}$, close to $20$, only within small regions of parameter space, with $\alpha \lesssim 2.1$ and $z_\text{CR}^{\max} \approx 0.03$. For $1 \leq \Gamma_\nu/\Gamma_\text{CR} \lesssim 11$, under the IC40 visibility criterion, the spectral index can take on values within the interval $2 \le \alpha \lesssim 2.65$, though the highest values are accessible only with $z_\text{CR}^{\max} = 0.01$ to $0.03$. For low values of $\Gamma_\nu/\Gamma_\text{CR}$, around $1$, the allowed ranges are $2.25 \lesssim \alpha \lesssim 2.65$ and $10^{-3} \le z_\text{CR}^{\max} \le 0.03$.

Using combined cosmic-ray data \cite{Berezinsky:2002nc}, the preferred value of $\alpha$ has been set at $2.7$. We have found that, if the AMANDA upper bound is used, this value is allowed in both the KT and BB models, whereas if the more recent IC40 upper bound is used, it is not. The authors of \cite{Becker:2008nf} claim that the true value of the spectral index might be either $\alpha = 2.0$ or $2.3$. For the BB model, these two values are allowed under both visibility criteria.  For the KT model, using the AMANDA bound, the value $\alpha = 2.3$ is allowed under strong source evolution, while under no source evolution it is not testable since it lies below the IC86 discovery potential. Using the IC40 bound, $\alpha = 2.3$ is excluded under strong source evolution and is also not testable under no source evolution. The value $\alpha = 2.0$ is not testable under any assumption on the source evolution. Note, however, that the experimental discovery potential and upper bounds that we have used were calculated for a $E_\nu^{-2}$ flux and that using them to constrain the BB and KT models might be slightly over-constraining the parameter space.

Additionally, in the event that an UHE neutrino signal is detected after five years of running the full IceCube array, and assuming that it was produced solely by the neutrino flux from AGN, we have explored the detector's capability to distinguish between the KT model, with strong and no source evolution, and the BB model, i.e., to determine which one of the two models would correctly describe the detected UHE neutrino data. In order to do this, we have defined a measure of the separation between the models as $\Delta\left(\alpha,\Gamma_\nu/\Gamma_\text{CR},z_\text{CR}^{\max}\right) \equiv \left\vert N_\text{BB}\left(\alpha,\Gamma_\nu/\Gamma_\text{CR},z_\text{CR}^{\max}\right) - N_\text{KT}\left(\alpha\right) \right\vert$, with $N_\text{BB}$ and $N_\text{KT}$ the number of muon-neutrinos expected in IceCube-86 associated to each model, between $10^5$ and $10^8$ GeV, after five years of running. At each point in parameter space, we have calculated the value of $\Delta$, expressed in units of $\sigma\left(\alpha\right) \equiv \sqrt{N_\text{KT}\left(\alpha\right)}$. The comparison between the flux models, however, is meaningful only in those regions of parameter space where both models simultaneously lie inside their respective visibility regions. Thus, under the IC40 visibility criterion, comparison is allowed only inside very small regions of simultaneous visibility located at $\alpha \simeq 2.57 (2.25) \-- 2.59 (2.27)$, $\Gamma_\nu/\Gamma_\text{CR} \simeq 1 (3) \-- 1.5 (4)$, and $z_\text{CR}^{\max} = 0.01 \-- 0.03$ assuming no (strong) source evolution. Within these regions, the separation between models is at the level of $5\sigma$ or higher. Hence, comparison between the models becomes unfeasible in most of the parameter space, but where it becomes possible, it is statistically meaningful.

A comment is in order: if, for the BB model, we had performed the integration in $z_\text{CR}^{\max}$ up to a value $\leq 5$, the associated number of events would have been larger and the corresponding visibility region even tighter than the ones we have presented, for which the contributions to the diffuse flux only come from the supergalactic plane ($z_\text{CR}^{\max} \leq 0.03$). Since the magnitude of the separation between models relies on the number of events, then either the level of separation would have been higher or there would have been no region of simultaneous visibility.

We have thus shown that, after five years of running, the completed IceCube array might be able to strongly constrain the KT and BB models, leaving only small regions of parameter space where the models survive. In addition, discrimination between the models, while feasible only within even smaller regions of parameter space, might be able to reach the $5\sigma$ level. The reader should be aware that our predictions are based on an all-proton cosmic-ray flux, but there is growing evidence that the UHECR flux is composed mainly of heavy nuclei \cite{Abraham:2009dsa,Hooper:2009fd,Abraham:2010yv,Abraham:2010mj} (see, however, \cite{Abbasi:2009nf,Wilk:2010iz}), and, as a consequence, the UHE neutrino flux would be reduced. Thus, with reservations, our results might be seen as symptoms of the need for new models of AGN neutrino production that are better equipped to face the latest experimental bounds on the UHE neutrino flux.

\acknowledgments

The authors would like to thank Julia Becker and Peter Biermann for helpful discussion of their neutrino production model; Hylke Koers and Peter Tinyakov for facilitating the neutrino boost factor calculated at the updated threshold energy; Kumiko Kotera, Teresa Montaruli and Sean Grullon for providing the estimates of the 40- and 86-string IceCube upper bound and discovery potential that we have used; and Jos\'e Luis Bazo for clarifying discussion. They would also like to thank the Direcci\'on de Inform\'atica Acad\'emica at the Pontificia Universidad Cat\'olica del Per\'u (PUCP) for providing distributed computing support through the LEGION system and Edith Castillo for her collaboration in the early stages of the work. This work was supported by grants from the Direcci\'on Acad\'emica de Investigaci\'on at PUCP through projects DAI-4075 and DAI-L009.

\appendix

\section{Neutrino detection in IceCube}\label{Section_App_IceCube}

We have calculated the predicted number of muon-neutrinos detected in IceCube-86 using the method presented in Ref.~\cite{Koers:2008hv}. In general, the integrated number of upgoing muon-neutrinos at a \u{C}erenkov detector due to a diffuse flux of muon-neutrinos, $\phi_{\nu_\mu}^\text{diff}$, with energies between $E_\nu^{\min}$ and $E_\nu^{\max}$, is calculated as
\begin{equation}\label{EqNumEvtsUpDn}
 N_{\nu,\text{up}} = T \Omega \int_{E_\nu^{\min}}^{E_\nu^{\max}} dE_\nu~ \phi_{\nu_\mu}^\text{diff}\left(E_\nu\right) A_{\nu,\text{eff}}^\text{up}\left(E_\nu\right) ~,
\end{equation}
where $T$ is the detector's exposure time; $\Omega$, the detector's opening solid angle; $E_\nu$, the neutrino energy; $\phi_{\nu_\mu}^\text{diff}$ is either the KT or BB diffuse AGN neutrino flux; and $A_{\nu,\text{eff}}^\text{up}$ is the upgoing neutrino effective area. 

Note that the six extra DeepCore strings of the IceCube-86 array increase the neutrino effective area only in the range $10 \le E_\nu/\text{GeV} \le 10^3$ \cite{Wiebusch:2009jf}. Above $10^3$ GeV, the IceCube effective area is determined solely by the remaining 80 strings.

The effective neutrino area takes the form
\begin{equation}\label{EqANuUp}
 A_{\nu,\text{eff}}^\text{up}\left(E_\nu\right) = S\left(E_\nu\right) P_\mu\left(E_\mu\right) A_{\mu,\text{eff}}\left(E_\mu\right) ~,
\end{equation}
where $S$ is the shadowing factor, which takes into account neutrino interactions within the Earth; $P_\mu$, the probability that the neutrino-spawned muon reaches the detector with energy greater than the threshold energy $E_\mu^{\min}$ required to be detected; and $A_{\mu,\text{eff}}$, the detector's effective area for muons. We will explain each term in eq.~(\ref{EqANuUp}) in what follows. 

\begin{figure}[t!]
 \begin{center}
  \scalebox{0.45}{\includegraphics{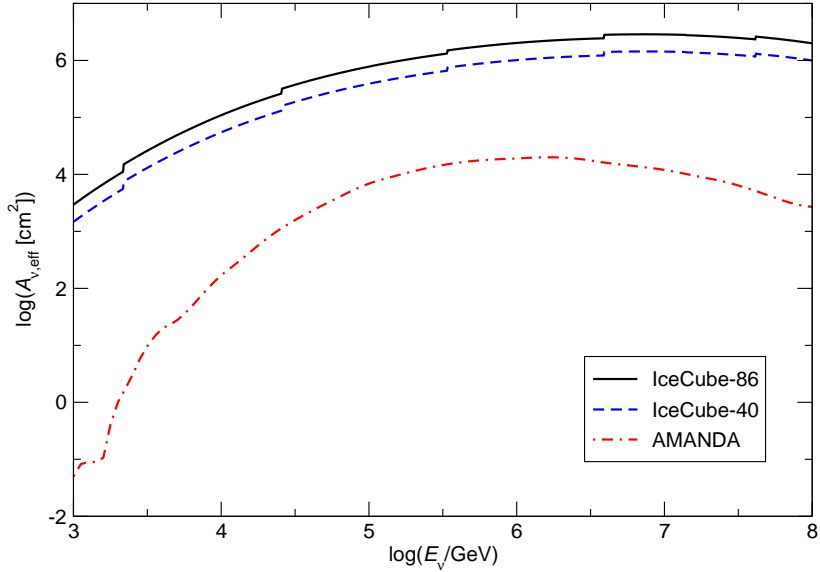}} 
  \caption{Angle-averaged upgoing neutrino effective areas, as functions of the neutrino energy, for the IceCube-86, IceCube-40, and AMANDA detectors. The IceCube-40 effective area is estimated at half the IceCube-86 area (see text), while the AMANDA effective area is a factor of $100$ lower than the IceCube-86 area.\label{Fig.Aeffnu}}
 \end{center}
\end{figure}

The probability of muon detection can be written as \cite{Koers:2008hv}
\begin{equation}
P_\mu\left(E_\mu\right) = 1 - \exp(-N_{\text{Av}} \sigma_{\nu N}^{\text{CC}}\left(E_\nu\right) R_\mu\left(E_\mu\right)) ~,
\end{equation}
where $N_{\text{Av}}=6.022\times10^{23}~\text{mol}^{-1} = 6.022\times10^{23}~\text{cm}^{-3}$ (w.e., water equivalent) is Avogadro's constant; $\sigma_{\nu N}^{\text{CC}}$ is the charged-current neutrino-nucleon cross section, taken from \cite{Gandhi:1998ri} (which uses CTEQ4 data); and $R_\mu$ is the muon range within which the muon energy reaches the threshold energy $E_\mu^{\min} = 100$ GeV, which can be expressed as
\begin{equation}
R_\mu\left(E_\mu\right) = \frac{1}{b}\ln\left(\frac{a+bE_\mu}{a+b E_\mu^{\min}}\right) ~,
\end{equation}
with $a=2.0\times10^{-3}$ GeV cm$^{-1}$ (w.e.) accounting for ionisation losses and $b=3.9\times10^{-6}$ cm$^{-1}$ (w.e.) accounting for radiation losses. The relation between neutrino and muon energy is obtained by assuming single-muon production in each neutrino interaction, which leads to $E_\mu = y_\text{CC}\left(E_\nu\right)E_\nu$, with $y_\text{CC}$ the mean charged-current inelasticity parameter tabulated in \cite{Gandhi:1998ri}.

The shadowing factor, $S$, is defined in terms of $P_\nu\left(E_\nu,\theta\right)$, the probability that a neutrino arriving at Earth with nadir angle $\theta$ (the North Pole is located at $\theta=0^\circ$) and interacting with Earth matter, reaches the detector. We use \cite{Koers:2008hv}
\begin{equation}
S\left(E_\nu\right) = \frac{1}{1-\cos\left(\theta_{\max}\right)} 
\int_0^{\theta_{\max}} d\theta \sin{\left(\theta\right)} P_\nu\left(E_\nu,\theta\right) ~,
\end{equation}
where $\theta_{\max}$ is the detector's maximum viewing angle, which we have taken to be $\theta_{\max}=85^\circ$, as in Ref.~\cite{Koers:2008hv}. Thus, the detector's opening angle is
\begin{equation}
\Omega 
= \int_0^{2\pi} d\phi \int_0^{\theta_{\max}} \sin\left(\theta\right) d\theta
\nonumber
= 2\pi\left[1-\cos\left(\theta_{\max}\right)\right]
\approx 5.736~\text{sr} ~.
\end{equation}
The neutrino survival probability can be written as
\begin{equation}
P_\nu\left(E_\nu,\theta\right) 
= \exp\left( -N_{\text{Av}} \sigma_{\nu N}^{\text{tot}}\left(E_\nu\right) \int_0^{L\left(\theta\right)} \rho\left(r\right) dl \right) ~,
\end{equation}
where $\sigma_{\nu N}^{\text{tot}}$ is the total (charged- plus neutral-current) neutrino-nucleon cross section, tabulated in Ref.~\cite{Gandhi:1998ri}; $\rho\left(r\right)$ is the Earth's density profile given by the Preliminary Reference Earth Model \cite{Gandhi:1995tf}, parametrised by the radial coordinate $r=\sqrt{l^2+r_E^2-2lr_E\cos\left(\theta\right)}$, with $r_E = 6371$ km the Earth radius; and $L\left(\theta\right) = 2r_E\cos\left(\theta\right)$ is the distance that a neutrino traversing the Earth at angle $\theta$ propagates.

Lastly, for IceCube-86's upgoing muon effective area, $A_{\mu,\text{eff}}$, we have used the curve corresponding to level-2 cuts in Figure 5 of Ref.~\cite{Ahrens:2003ix}, which is the effective area averaged over the northern hemisphere, and dependent only on the incoming muon energy, $E_\mu$. Figure \ref{Fig.Aeffnu} shows that the IceCube-40 neutrino effective area is estimated at one half the IceCube-86 effective area (see Section \ref{Section_Bounds}), while the AMANDA neutrino effective area was a factor of $100$ lower than IceCube-86 area.

\end{document}